\documentstyle[11pt,paspconf]{article}

\begin{document}

\title{Overview of Grain Models}
\author{Adolf N. Witt}
\affil{Ritter Astrophysical Research Center, The University of Toledo, 
    Toledo, OH 43606, USA}

\begin{abstract}

The properties of interstellar grains can now be defined by a rapidly growing wealth of observational data. We rely upon models to combine these data with unobserved properties such as the size distribution of grains, their structure and shape, and their detailed chemical composition, in a self-consistent manner.
I will review the observational constraints which define the boundaries for possible models and will then summarize the features of the three quasi-comprehensive grain models most commonly referred to in the current literature. I will conclude by discussing recent work dealing with single aspects of interstellar grains and the phenomena attributed to them.
\end{abstract}

\keywords{interstellar: grains, absorption features, emission features}

\section{Introduction}

A comprehensive model for interstellar grains should provide detailed 
information on the composition, size distribution, optical properties, 
and physical structure and shape of interstellar grains, fully consistent with the rapidly expanding list of observational constraints, including extinction,
scattering, polarization, and emission properties, spectroscopic absorption
and emission features, and the abundances of refractory elements in the 
interstellar medium (ISM) and their observed depletion pattern. In addition,
such a model should also have predictive capabilities, allowing one to 
correctly anticipate observational phenomena not yet seen. Despite substantial
progress in the nearly seven decades since the recognition of interstellar dust
as a general astrophysical phenomenon (Trumpler 1930), such a comprehensive
dust model does not yet exist. Instead, we have several quasi-comprehensive
models which satisfy at least some sub-set of observational constraints, and
an even larger number of more limited, constraint-specific models, which 
have been designed to explain single observational constraints without
even attempting to approach comprehensiveness. In addition, much theoretical
work has been directed at understanding the sources and processes of grain
formation and the mechanisms involving grain-grain and grain-gas interactions
in the interstellar environment, which again must be consistent with the
observed gas-to-dust ratio in interstellar clouds and the spectroscopic
signatures of various grain components.

A study of the model-to-model differences provides a vivid illustration of the fundamental issues concerning interstellar grains, about which there remains
substantial disagreement. These issues include uncertainties about the details
of the size distribution, especially near the upper and lower ends of the distribution, questions about grain composition, especially the amount, nature, and structure of carbon-based grains, and disagreements about the physical grain structure.

During the past decade, several comprehensive reviews of the interstellar
dust problem have appeared in print, foremost among these the monograph by
Whittet (1992) and the reviews by Mathis (1993) and by Dorschner and 
Henning (1995). The interested reader is advised to go to these sources for
additional background information. The present review will focus entirely upon dust in the diffuse interstellar medium of the Milky Way galaxy, and it will aim to summarize the current status of observational constraints with emphasis on more recent developments, followed by a discussion of the principal semi-comprehensive models and their abilities to match these constraints. I will then review important recent work and theoretical explorations related to dust models, which will highlight the impressive rate at which new observational advances help define the detailed properties and characteristics of interstellar
grains.

\section{Observational Constraints}

\subsection{Chemical Abundances and Depletions}

The absence of interstellar absorption features attributable to cosmic ices along interstellar lines of sight traversing only the diffuse ISM indicates that
the dust in this environment is composed only of refractory solids, consisting mainly  of the relatively abundant heavy elements C, O, Fe, Si, Mg, S, Ca, Al, and Ni (Whittet 1992). The abundance of these elements relative to hydrogen in the current interstellar medium and the degree to which these elements are depleted from the gas phase determine the overall chemical composition of dust grains and their combined mass relative to the interstellar gas. Until quite recently, it was accepted practice to equate the interstellar relative abundance of heavy elements with that encountered in the Sun and the solar system. Recent evidence, based on abundance determinations in young stars, i.e. objects formed more recently from the local ISM, on abundance determinations in HII regions,  on studies of the distribution of the heavy-element abundance in solar-type stars at the Sun's galactocentric distance, and on abundance determinations of
undepleted elements in the diffuse ISM itself
has led to suggestions that the actual abundances of heavy elements in the present Galactic ISM , the so-called reference abundances, are lower than the
earlier assumed solar abundances, possibly by about 1/3 of the total (Snow \& Witt 1996). Combined with new observations of the gas-phase abundances along
diffuse ISM lines-of-sight (e.g. Fitzpatrick 1996; Sofia et al. 1997; Meyer et al. 1998) obtained with the GHRS on HST, tight new constraints on the chemical composition on interstellar grains, especially on carbonaceous grains (Snow \& Witt 1995; Meyer 1997), have emerged. All current dust models have some difficulty in meeting these constraints, both with regard to the carbon abundance and with regard to matching the implied interstellar mass extinction coefficient for interstellar grains.

\subsection{Interstellar Extinction}

Extinction by dust in interstellar space adds two important constraints which
help to specify the nature of grains. The wavelength dependence of extinction,
now known over a wavelength interval extending from 0.1 $\mu$m to $\sim$1000
$\mu$m, provides important information about the size distribution of
dust particles. The total amount of optical extinction for a sightline with known hydrogen column density is the prime determinant for the interstellar dust-to-gas ratio.
Observed reddening throughout the near-IR and optical wavelength ranges restricts the bulk of the interstellar dust to the sub-$\mu$m size range.
The continued non-linear rise of extinction throughtout the far-UV points to the existence of grains small compared to UV wavelengths. The relative
amounts of extinction in the visual and at far-UV wavelengths helps to specify the mass ratio of very small grains absorbing mainly in the UV to the larger
sub-$\mu$m grains contributing absorption and scattering throughout the 
near-IR/optical/UV spectral range. The large degree of spatial variation in
this ratio (Fitzpatrick 1999) indicates that variations in the size distribution
are similarly large within our Galaxy.

\subsection{Scattering by Dust}

Observations of scattering by dust can provide useful information about the
nature of grains and their size distribution by yielding data on the albedo and the phase function asymmetry and their respective wavelength dependences.
The high degree of linear polarization of scattered light can provide additional insight, provided the dust/source geometry is sufficiently well known.
Finally, the scattering of x-rays by interstellar dust, both through the 
intensity level of scattered x-rays and the radial distribution of intensity in the resulting x-ray halos, constrains the composition and structure as well as the size distribution of interstellar grains, especially at the large-size end of this distribution.
Examples of particularly useful scattering results are of the reflection nebula NGC 7023 in the 160 to 100 nm wavelength range (Witt et al. 1993) which
show that the far-UV albedo declines shortward of 130 nm, suggesting that the
particles responsible for the far-UV rise in extinction are mainly small absorbing grains. Similarly, studies of the nebula IC 435 (Calzetti et al. 1995)
demonstrate rather definitively that the 2175 \AA\ extinction ``bump'' is caused entirely by absorption and that the asymmetry of the scattering phase function increases monotonically from the visual into the far-UV range. The latter finding shows clearly that scattering throughout the visible and UV is caused by the same relatively large grains, which produce very strongly forward-directed scattering in these wavelength ranges.
The potential of using dust scattered x-ray haloes as a grain diagnostic has been reviewed recently by Predehl and Klose (1996), and examples of the analysis of the particularly well-observed x-ray halo of Nova Cygni 1992 are the papers by Mathis et al. (1995) and by Smith and Dwek (1998).

\subsection{Interstellar Polarization by Aligned Grains}

The wavelength dependence of partial linear and circular polarization (Martin 1989) caused by interstellar dust in the diffuse ISM reveals unique information on the shape and size distribution of alignable grains and the mechanism 
producing this alignment. The efficiency of polarization, i.e. its degree relative to the associated amount of visual extinction, provides a measure for the effectiveness of this alignment process. An important expansion of the polarization data base has occurred recently with the addition of measurements of ultraviolet linear interstellar polarization (Martin et al. 1999).

\subsection{Discrete Interstellar Absorption Features}

Discrete spectroscopic absorption  features offer the potential of identifying specific material components of grains. Several such features are seen in the diffuse ISM. The strongest is the 2175 \AA\ UV extinction feature, which most models attribute to absorption by carbonaceous, most likely graphitic, small
grains (Draine 1989). Still less controversial is the identification of the infrared interstellar absorption bands at 9.7 and 18 $\mu$m with the Si-O 
stretch and Si-O-Si bending modes in amorphous silicates and the attribution of
the infrared interstellar absorption feature at 3.4 $\mu$m to C-H stretch transition in aliphatic hydrocarbon solids. On the other hand, the identification of the numerous diffuse interstellar absorption bands (Sarre 1999, this volume) remains one of the most challenging unsolved problems of interstellar spectroscopy. Until these bands are positively identified, their great potential as diagnostics of physical conditions in the ISM will not be realized. Equally poorly understood is the very-broadband structure (VBS) in 
the visual interstellar extinction curve (Hayes et al. 1973). The VBS takes the form of a shallow depression in the extinction in the 520 - 600 nm region, 
seen in many Galactic lines of sight (e.g. van Breda \& Whittet 1981). Efforts to connect the VSB to the occurrence of photoluminescence by grains (Duley \& Whittet 1990) or to UV-characteristics of interstellar extinction (Jenniskens 1994) have not led to a conclusive identification of the VSB carrier.

\subsection{Thermal Emissions from Dust}

Interstellar grains are heated by absorption of UV/optical photons and they re-radiate this energy as thermal emission with a spectrum revealing the temperature distribution and emission characteristics of the radiating particles. The near-full extent of this emission, both spatially and in wavelength (Beichman 1987; Soifer, Houck, and Neugebauer 1987), was first seen with the Infrared Astronomical Satellite (IRAS), then further expanded with the
Diffuse Infrared Background Explorer (DIRBE) (Bernard et al. 1994). 
In addition to the expected radiation from relativlely cool (approx. 20K) wavelength-sized grains in equilibrium with the local radiation field, IRAS
revealed the presence of non-equilibrium thermal emission from small and very small grains, i.e. nanoparticles, whose heat capacity is so small that the absorption of a single UV/optical photon results in large (500 - 1000K)
temperature excursions. With about 35\%\ of the entire diffuse IR radiation 
from the Galaxy originating with this process, the important role of nanoparticles in the size distribution of interstellar grains has now been
fully recognized (Puget \& Leger 1989). 

\subsection{Discrete Interstellar Emission Features}

Emission features originating in interstellar grains are observed in the 
optical in the form of Extended Red Emission (ERE) and in the near-infrared 
spectral range in the form of Aromatic Hydrocarbon Bands, also referred to
as the Unidenditified Infrared Bands (UIB). The ERE was first seen in the peculiar reflection nebula called the Red Rectangle (Schmidt et al. 1980),
where it appears as an exceptionally intense, broad emission feature extending from 540 nm to about 900 nm with a peak near 660 nm. Subsequently, the ERE 
feature has been observed in many other dusty environments, including reflection
nebulae, planetary nebulae, HII-regions, external galaxies, and the diffuse ISM
of the Milky Way (Gordon, et al. 1998; Szomoru \& Guhathakurta 1998). The ERE is
an efficient photoluminescence process associated with interstellar grains,
as shown by its close correlations with the dust column density and the intensity of the illuminating radiation field. In the diffuse ISM, at least 10\%
of the interstellar UV/optical photons are absorbed by the ERE carrier (Gordon,
et al. 1998). This calls for a carrier consisting of cosmically abundant elements, and current ideas center on silicon nanoparticles (Witt et al. 1998;
Ledoux et al. 1998; Zubko et al. 1999), carbonaceous nanoparticles (Seahra \& Duley 1999), and PAH molecules (d'Hendecourt et al. 1986).
The UIBs, with principal emission features centered at wavelengths of 3.3, 6.2,
7.7, 8.6, and 11.3 $\mu$m, after having been observed mainly in dense dusty environments with abundant UV photons for many years, have now also been detected as prominent features in the diffuse ISM (Giard et al. 1989; Mattila et al. 1996; Onaka et al. 1996) through obvervations from the balloon experiment
Arome and the infrared satellites ISO and IRTS. Several current dust models (see below) attribute the UIBs to emission from PAH molecules, although solid-state carbonaceous models have been advanced as well (Papoular et al. 1996; Duley \& Williams 1981). 

\subsection{Interstellar Grains in the Solar System}

Evidence of interstellar grains is found in the solar system in two ways: either in the form of pre-solar grains extracted from primitive meteorites (Zinner 1997) and liberated from evaporating comets (Bradley et al. 1999) or in the form
of present-day interstellar grains entering the solar system from interstellar space (Landgraf et al. 1999). Isotopic anomalies in presolar grains provide unique insights into the sources of certain sub-populations of grains, be they
asymptotic giant branch stars or supernovae; they provide evidence that crystalline grains can form in astronomical environments and survive. A population of
glassy silicate, interplanetary dust particles, known as GEMS, exhibit  9.7 $\mu$m silicate features closely matching those of silicates seen in astronomical sources (Bradley et al. 1999). The possibility that GEMS are examples of actual
interstellar silicate grains must therefore be considered seriously.
Dust detectors on space craft exploring the outer reaches of the solar system 
are detecting interstellar grains entering interplanetary space from the local interstellar medium (Landgraf et al. 1999). The analytical capabilities of these detectors are still limited, restricting the measurements to that of mass flux and velocity vector. Nevertheless, the data available have provided new information on the upper end of the dust size spectrum and the dust-to-gas ratio
applicable at least to the interstellar grains in the local interstellar cloud
(Frisch et al. 1999). The successful launch of STARDUST in 1999 provides hope that actual samples of cometary and interstellar dust will be captured and returned to Earth within only a few years.

\section{Major Dust Models in the Current Literature}

\subsection{The MRN Model}

Mathis, Rumpl, and Norsieck (1977), with a focus on explaining the entire
observed wavelength dependence of interstellar extinction and using as a constraint the relative abundances of refractory elements based on solar
reference abundances, proposed a grain model consisting of separate spherical
particles consisting of graphite and silicates. This model is known as the 
MRN model. The proposed grain compositions take into account that there are
two basic chemical environmemts for the condensation of solids from the gas
phase, depending upon the C/O ratio being less than or greater than unity.
In the first case, all C-atoms end up in CO, with the remaining O-atoms
available to form silicates and metal oxides. In the second case, all O-atoms end up in CO, with the remaining C-atoms free to form various carbonaceous
compounds, although not necessarily graphite. The model ignores the likely processing experienced by interstellar grains through exposure to shocks, UV radiation, cosmic rays, and cold, dense gas in molecular clouds.
A defining characteristic of the MRN model is its assumed power-law size
distribution n(a) $\sim a^{-3.5}$, with sharp cutoffs at both a maximum radius (250 nm) and a minimum radius (5 nm). This model lacked  components which
would account for the observation of widespread UIB and ERE emissions in the
diffuse ISM; it does not account for the 3.4 $\mu$m interstellar absorption, and, lacking nanoparticles, it did not anticipate the important role of 
non-equilibrium thermal emission, which is an important part of the IR spectrum
of the Milky Way  and of other dusty galaxies.
The MRN model received major upgrades through the work of Draine \& Lee (1984) and Draine \& Anderson (1985), who, respectively, provided the optical constants
for ``astronomical silicates'' and ``astronomical graphites'' to provide a more satisfactory fit to the interstellar extinction curve, and who added a population of ``very small grains'', filling the size range between 5 nm
(the previous MRN lower limit) and 0.3 nm, the size of large molecules.
The latter adjustment allowed the MRN model to account for the existence of
the near-IR continuum emission, albeit not for the UIB structure of this emission. While long abandoned by its principal originators, the MRN model continues to be most frequently invoked when references to interstellar grains
are being made. It is clearly not the comprehensive model called for in the introduction.

\subsection{Core-Mantle Models}

Core-mantle models are based on the idea that the bulk of the carbonacous grain component resides as a mantle on the surfaces of silicate grains. Greenberg and his collaborators (Greenberg \& Hong 1974; Hong \& Greenberg 1980; Li \& Greenberg 1997) envision this carbonaceous mantle in the form of an organic
refractory residue, which is produced via photolysis of icy grain mantles formed on grain surfaces while interstellar grains spend time periodically inside dense molecular clouds. In its latest version (Li \& Greenberg 1997), this model assumes a trimodal size distribution consisting of large core-mantle grains containing most of the mass, a second population of small carbonaceous grains
of graphitic nature to produce the 2175 \AA\ feature, and a population of PAH molecules to provide the rising far-UV extinction through absorption and the source for the UIB emission. This model provides a satisfactory match to the interstellar extinction curve, and, through the large core-mantle grains, of the interstellar polarization. It also give a satisfactory fit to measured scattering properties such as dust albedo and phase function asymmetry, and most importantly, it fits the infrared dust emission spectrum. At the same time, it comes relatively close to fitting current interstellar chemical abundance constraints.

Weaknesses of this model are the poor characterization of the population of small carbonaceous grains and the still open question of how the core-mantle structure can be maintained through the shattering experience of encountering an interstellar shock. If processes like grain shattering and subsequent reassembly through agglomeration of grain fragments are important in the life cycles of grains, a more randomly composite grain structure as envisioned by Mathis \& Whiffen (1989) appears more likely.

Another version of core-mantle grain models has been proposed by Duley, Jones, \& Williams (1989) and Jones, Duley, \& Williams (1990). They suggest that the
carbonaceous mantle is in the form of hydrogenated amorphous carbon (HAC), which can experience modifications of its optical properties by changing the degree of
hydrogenation in  different interstellar environments (Duley \& Williams 1990).
This was intended to provide explanations both for the high degree of spatial variation observed in the far-UV extinction curve plus for the widely seen ERE,
for which HAC was considered a possible source material. It should be recognized that the type of evolutionary processes envisioned by both types of core-mantle grain models most likely do occur at some level.

\subsection{The Post-IRAS Model}

Building upon the lessons learned from IRAS, Desert, Boulanger, \& Puget (1990)
proposed a dust model with a focus on explaining the energetics and spectrum of near-IR, mid-IR, and far-IR dust emission and the implied very important role of very small grains and PAH molecules. They also suggest three independent grain populations with radii ranging from 110 nm to 15 nm (big grains), 15 nm to 1.2 nm (very small grains), and 1.2 nm to 0.4 nm (PAHs). They envision relatively large spatial abundance variations among these three populations to account for the large variations seen in the far-UV extinction as well as in the details of the IR dust emission spectrum. This model can claim that it correctly predicted
the UIB emission confirmed later by the Arome, IRTS, and ISO experiments.
A weakness of this model is the small maximum size of the ``big grains''.
They may prove to be insufficient to explain interstellar linear polarization
observations, observations of near-IR scattering, and observations of x-ray halos, and observations of present-day interstellar grains found entering the solar system.

\subsection{Common Properties and Problems}

While differing in details, all these models have certain properties in common as they also share certain problems. All agree that silicates are a major component of the large-grain population, playing the most important role in visual extinction and interstellar reddening, in scattering at all wavelengths,
in interstellar linear polarization, and in the far-IR dust emission at wavelengths around 100 $\mu$m and beyond. The observed strength of the 
``silicate bands'' at 9.7 and 18 $\mu$m requires the near total depletion of gas-phase Si of solar abundance and its incorporation in Si-O bonds in silicates. The models also agree that size distribution of grains must extend in a near-continuous fashion from the sub-$\mu$m range to the nanoparticle range,
forming a continuous transition to large molecules. The number density of grains
must increase with decreasing size in a manner of a power law with negative power of 3 or larger. All models have the problem of meeting the interstellar
chemical abundance constraints as currently understood, especially the limits
on the Si- and C-abundances. There is considerable diagreement on the dominant form and structure of carbonaceous dust.

\section{Recent Work Related to Dust Models}

Spurred on by the very impressive growth in the amount of observational information on dust-related phenomena, research directed at solving some of the outstanding problems of interstellar grain models is progressing at an ever-increasing pace. This is aided by the availability of increasingly sophisticated laboratory facilities and of increasingly powerful computing 
resources and numerical techniques. Space limitations prevent me from 
commenting on laboratory studies in any detail. These provide absolutely essential data on optical properties of relevant solids, on wavelengths, profiles, and absorption strengths of absorption bands, on the emission characteristics of proposed carriers of the UIBs and the ERE, and on the processing of likely grain analog materials by exposure to radiation and to gases. Other efforts related directly to recent grain modeling work will be reviewed briefly below.

\subsection{New Tighter Abundance Constraints}

Realizing that the relative abundances of refractory elements in the interstellar medium  may be less than expected on the basis of their abundances in the Sun and the solar system (Snow \& Witt 1995, 1996; Meyer 1997), Mathis (1996) began the careful examination of grain models under these new, tight abundance constraints. Fluffy grains, having a porous structure with up to
50\% vacuum, were found to hold out promise for a sufficiently increased mass extinction coefficient so that the new constraints might be met, albeit only barely within the still large observational uncertainties. This possibility
had already been suggested earlier by Jones (1988) in a different context. Dwek (1997) pointed out, however, that the pourous-grain model predicts a far-IR emissivity in excess of that observed from the diffuse ISM with the Cosmic Background Explorer (COBE), resulting from the lower dust albedo of the fluffy grains compared to traditional models and direct observations. Such problems should be surmountable, if the modeling approach includes the abundance constraints along with other observational constraints in defining the composition and size distribution of interstellar grains, as has been shown by Zubko et al. (1998) using the regularization approach (Zubko 1999) as a base for developing a modern dust model.

\subsection{Grains with Composite-Porous-Fluffy-Irregular Structure}

For many decades, the Mie-theory (see for reference: Bohren \& Huffman 1983)
has been the standard tool for computing the optical properties of model grains,
given a set of indices of refraction from laboratory measurements for the assumed grain material. This, however, limited models to spherical, homogeneous, optically isotropic particles, spherically concentric core-mantle particles, or infinite cylinders in as far as sub-$\mu$m grains were included.
Purcell \& Pennypacker (1973), followed by Draine (1988), introduced the
``discrete dipole approximation'' (DDA) as a practical, albeit computing-intensive, technique for calculating the optical properties of irregularly shaped, composite, and/or optically anisotropic wavelength-sized grains. Coupled with a faster iterative method introduced by Lumme \& Rahola
(1994), the DDA appears to be the method of choice in the age of rapidly increasing computing power, although optical properties of composite particles
which do not deviate too much from spherical shape may be computed quite reliably via classical Mie theory with optical constants derived from an effective medium theory. One of the latest published efforts of modeling extinction and infrared emission assuming fractal dust grains is the work of
Fogel \& Leung (1998). Among their results, their finding that interstellar
extinction with fractal grains requires about one-third less mass in the form of grains is most interesting, as this is just the fraction of mass having been made unavailable through the tighter abundance constraints.

\subsection{Grain Size Distributions}

Every one of the three quasi-comprehensive dust models reviewed in Section 3
uses a different ad-hoc form for the size distribution of grains, with several free parameters to be determined by fitting the model predictions to the observations. In all three cases, the largest grains contain most of the dust mass available. The upper size limit is thus set by the dust-to-gas ratio derived from observations  and the desire to avoid grey extinction which does not contribute to interstellar reddening. The lower size limit, initially, was
ill-defined by extinction constraints, but it is now well-determined by the requirement of non-equilibrium thermal emission, resulting from the stochastic heating of very small grains by individual photons, and it lies below 1 nm.
Two techniques have recently been advanced for the determination of size distributions that aim to find the optimum distribution derived from observational constraints alone. The maximum entropy method (MEM) was used by
Kim et al. (1994) and Kim \& Martin (1994, 1995) toward this end, and although they assumed bare silicate and graphite spherical grains consistent with the MRN model, they found significant departures from the simple power law assumed by MRN. In particular, they found the sharp upper size limit to be quite unrealistic and saw their MEM distribution extend toward larger grain sizes.
A direct result of this was their prediction that near-IR dust albedos should be higher by about a factor two compared to MRN, which appears to be consistent with observations. 

While the MEM approach still requires an initial guess for the size distribution, the regularization technique employed by Zubko and his collaborators (Zubko 1999a; Zubko et al. 1998, 1999) computes the size distribution from the observational and abundance constraints alone. This opens the possibility of determining effective size distributions of grains for individual lines of sight (Zubko et al. 1996, 1998) and for extragalactic
systems such as the SMC (Zubko 1999b) where the extinction curve is known.

Given the impact of the grains near the upper limit of the size distribution on the dust-to-gas mass ratio, efforts to determine the shape of the distribution 
at this end by independent means are of particular importance. Ongoing in-situ  collections of interstellar grains by spacecraft in the outer solar system are especially significant.
Landgraf et al. (1999)  have summarized the results to-date, after showing that detected interstellar grains can be separated cleanly from a background of solar-system dust. Frisch et al. (1999) discuss the consequences of these findings on the implied dust-to-gas ratio in the local interstellar cloud.
Values typically twice those  generally accepted for the average diffuse ISM in the plane of the Milky Way galaxy are found, but given the local nature of the data, the implications are locally restricted. Results of more general validity
can be expected from the analysis of x-ray haloes, whose radial distribution
and absolute intensity for a given column density of grains is most strongly influenced by the largest grains in the line-of-sight size distribution
(Smith \& Dwek 1998). The lines-of-sight are typically of the order of kpc,
and with the advent of powerful x-ray telescopes such as Chandra and XXM,
suitable data are expected to be available in the near future.

\subsection{ The 2175 \AA\ UV Absorption Feature}

The current grain models explain the 2175 \AA\ feature through absorption by
graphite grains. This explanation is beset with numerous difficulties (Draine \&
Malhotra 1993; Mathis 1994; Rouleau et al. 1997). Finding a satisfactory alternative to graphite grains, ideally a type of grain which can explain other dust features in addition to the 2175 \AA\ absorption feature, has therefore been at the center of numerous investigations in recent years.

The difficulty of forming graphite grains in an initial condensation process
has been recognized for a long time (Czyzak et al. 1981). As a result, many 
studies have been directed at processes by which hydrogenated amorphous carbon grains, which are believed to be more easily formed in carbon star atmospheres, might be graphitized in interstellar space ( Sorrell 1990; Blanco et al. 1991, 1993; Mennella et al. 1995, 1997, 1998). UV-irradiation appears to be the most plausible process, and indeed it does lead to an absorption profile which
approaches that of the observed 2175 \AA\ band (Mennella et al. 1998). Hydrogenated amorphous carbon, in a more hydrogenated form, could then also be drawn upon as the source material for the interstellar 3.4 $\mu$m absorption band (Furton \& Witt 1999; Mennella et al. 1999)

Polycyclic aromatic hydrocarbons (PAH) are believed to be abundant in interstellar space and are considered a possible carrier of the UIB emission.
PAH exhibit particularly strong absorption in the UV. The old idea that a combination of numerous PAH absorption spectra might result in the observed
2175 \AA\ feature (Donn 1968) has been revived recently. Beegle et al. (1997)
found in laboratory experiments that molecular aggregates produced from the PAH naphtalene exhibit a 2175 \AA\ feature. Duley and Seahra (1998) investigated
the optical properties of carbon nanoparticles composed of stacks of PAHs, involving up to several hundred carbon atoms, and found that such structures
could indeed explain the plasmon-resonance type extinction feature at 2175 \AA.
In their model, these nanoparticles simply represent the high-mass component of
a general population of large PAH molecules in interstellar space. Experimental
work by Schnaiter et al. (1998) with nano-sized hydrogenated carbon particles
have shown promising results.

Other alternatives have been explored by Henrard et al. (1993) and Ugarte (1995)
along lines of onion-like graphitic particles or multishell fullerenes. Such particles might be generated through annealing of tiny nano-diamonds, which are found with remarkable abundance in primitive meteorites and which are demonstrably pre-solar (Zinner 1997). Papoular et al. (1993), within the context
of their coal model for carbonaceous interstellar grains, found that anthracite
produces a close fit to the 2175 \AA\ feature. 

\subsection{Extended Red Emission}

With the detection of intense extended red emission (ERE) in the diffuse ISM of the Milky Way galaxy ( Gordon et al. 1998;  Szomoru \& Guhathakurta 1998),
the ERE has become an important observational aspect of interstellar grains that future models need to reproduce. The observational evidence shows that ERE is a photoluminescence process powered by far-UV photons but that the ERE carriers are easily modified by intense UV radiation fields and destroyed, if the 
radiation intensity exceeds certain levels of intensity and hardness. This behavior points to large molecules or nanoparticles as the likely carrier.
Witt et al. (1998) and Ledoux et al. (1998) have proposed silicon nanoparticles
in the size range from 1 to 5 nm as the likely candidates, supported by an extensive base of laboratory data. Seahra \& Duley (1999) suggested that the
same PAH nanoparticles  proposed by them as the 2175 \AA\ band carriers
are also the source of the ERE. The number of interstellar photons absorbed in the 2175 \AA\ band is approximately what is required to excite the ERE (Gordon et al. 1998), but this does not necessarily imply a causal connection. The Seahra
\& Duley model predicts, in addition to the main ERE band, two satellite bands, one shortward of 500 nm and one at 1000 nm wavelength. No evidence for 
photoluminescence by grains shortward of 500 nm has been found, however (Rush \& Witt 1975).

\section{Open Questions}

Despite some very impressive progress in our understanding of interstellar grains, spurred on strongly by exciting new observations and laboratory results,
a number of challenging open questions remains. I will mention a few of my favorite ones. 

\begin{itemize} 

\item What are the dominant, most abundant forms of carbon solids in interstellar space?

\item How can carbon solids be sufficiently multi-functional to meet abundance constraints and yet explain all interstellar phenomena attributed to carbonaceous grains?

\item What is the nature of the  nanoparticles which radiate in the 25 to 60 $\mu$m range in the diffuse ISM?

\item What are the appropriate indices of refraction for grain materials, if they are composed of conglomerates of nanoparticles?

\item Why does silicon get in and out of grains so readily?

\item What produces the ERE?

\item Why do starburst galaxies seem to lack the 2175 \AA\ absorber?

\end{itemize}

\acknowledgments

I am grateful to the Scientific Organizing Committee for inviting me to attend
this symposium and to the local Korean  hosts for providing a beautiful setting for a week of stimulating interactions. I gratefully acknowledge financial support received from the Organizing Committee, from The University of Toledo, and from the National Aeronautics and Space Administration.


\begin{references}
\reference Beegle, L.W., Wdowiak, T.J., Robinson, M.S., Cronin, J.R., McGehee, M.D., Clemett, S.J., \& Gillette, S. 1997, \apj\ 487, 976
\reference Beichman, C.A. 1987, \araa\ 25, 521
\reference Bernard, J.P., Boulanger, F., Desert, F.X., Giard, M., \& Puget, J.L. 1994, \aap\ 291, L5
\reference Blanco, A., Bussoletti, E., Colangeli, L., Fonti, S., \& Stephens, J.R. 1991, \apj\ 382, L97
\reference Blanco, A., Bussoletti, E., Colangeli, L., Fonti, S., Mennella, V., \& Stephens, J.R. 1993, \apj\ 406, 739
\reference Bohren, C.F., \& Huffman, D.R. 1983, Absorption and Scattering of Light by Small Particles, (New York: John Wiley \& Sons), Chapt. 4
\reference Bradley, J.P. et al. 1999, SCIENCE 285, 1716
\reference Calzetti, D., Bohlin, R.C., Gordon, K.D., Witt, A.N., \& Bianchi, L. 
    \apj\ 446, L97
\reference Czyzak,S.J., Hirth, J.P., \& Tabak, R.G. 1981, Vistas in Astronomy 25, 337
\reference Desert, F.-X., Boulanger, F., \& Puget, J.L. 1990, \aap\ 237, 215
\reference d'Hendecourt, L.B., Leger, A., Olofsson, G., \& Schmidt, W. 1986, \aap\
170, 91
\reference Donn, B. 1968, \apj\ 152, L129
\reference Dorschner, J. \& Henning, T. 1995, The Astron. Astroph. Rev., 6, 271
\reference Draine, B.T. 1988, \apj\ 333, 848
\reference Draine, B.T. 1989, IAU Symp. 135, ed. L.J. Allamandola \& A.G.G.M. Tielens, (Dordrecht: Kluwer), p.313
\reference Draine, B.T., \& Anderson, N. 1985, \apj\ 292, 494
\reference Draine, B. T., \& Lee, H.M. 1984, \apj\ 285, 89
\reference Draine, B.T., \& Malhotra, S. 1993, \apj\ 414, 632
\reference Duley, W.W., Jones, A.P., \& Williams, D.A. 1989, MNRAS 236, 709
\reference Duley, W.W., \& Seahra, S. 1998, \apj\ 507, 874
\reference Duley, W.W., \& Whittet, D.C.B. 1990, MNRAS 242, 40P
\reference Duley, W.W., \& Williams, D.A. 1981, MNRAS 196, 269
\reference Duley, W.W., \& Williams, D.A. 1990, MNRAS 247, 647
\reference Dwek, E. 1997, \apj\ 484, 779
\reference Fitzpatrick, E.L. 1996, \apj\ 473, L55
\reference Fitzpatrick, E.L. 1999, \pasp\ 111, 63
\reference Fogel, M.E., \& Leung, C.M. 1998, \apj\ 501, 175
\reference Frisch, P.C. et al.  1999, \apj\ in press
\reference Furton, D.G., Laiho J.W. \& Witt, A.N. 1999, \apj\ in press
\reference Greenberg, J.M., \& Hong, S.S. 1974, Galactic Radio Astronomy, Proc. IAU Symp. No. 60, ed. F.J. Kerr and S.C. Simonson III, (Dordrecht: Reidel), p. 155
\reference Gordon, K.D., Witt, , A.N., \& Friedmann, B.C. 1998, \apj\ 498, 522
\reference Hayes, D.S., Mavko, G.E., Radick, R.R., Rex, K.H., \& Greenberg, J.M.
1973, Interstellar Dust and Related Topics, ed. J.M. Greenberg and H.C. van de Hulst, (Dordrecht: Reidel Publ. Co.), p.83
\reference Henrard, L., Lucas, A.A., \& Lambin, Ph. 1993, \apj\ 406, 92
\reference Hong, S.S., \& Greenberg, J.M. 1980, \aap\ 88, 194
\reference Jenniskens, P. 1994, \aap\ 284, 227
\reference Jones, A.P. 1988, MNRAS 234, 209
\reference Jones, A.P., Duley, W.W., \& Williams, D.A. 1990, QJRAS 31, 567
\reference Kim, S.-H., \& Martin, P.G. 1994, \apj\ 431, 783
\reference Kim, S.-H., \& Martin, P.G. 1995, \apj\ 444, 293
\reference Kim, S.-H., Martin, P.G., \& Hendry, P.D. 1994, \apj\ 422, 164
\reference Landgraf, M., Baggaley, W.J., Gruen, E., Krueger, H., \& Linkert, G. 1999, J. Geoph. Res. (in Press)
\reference Ledoux, G. et al. 1998, \aap\ 333, L39
\reference Li, A., \& Greenberg, J.M. 1997, \aap\ 323, 566
\reference Lumme, K., \& Rahola, J. 1994, \apj\ 425, 653
\reference Martin, P.G 1989, IAU Symp. 135, ed. L.J. Allamandola \& A.G.G.M. Tielens, (Dordrecht: Kluwer), p.55 
\reference Martin, P.G., Clayton, G.C., \& Wolff, M.J. 1999, \apj\ 510, 905
\reference Mathis, J.R. 1994, \apj\ 422, 176
\reference Mathis, J.S. 1996, \apj\ 472, 643
\reference Mathis, J.S. 1993, Rep. Progr. Phys., 56, 605
\reference Mathis, J.S., Cohen, D., Finley, J.P., \& Krauter, J. 1995, \apj\ 449, 320
\reference Mathis, J.S., Rumpl, W., \& Nordsieck, K.H. 1977, \apj\ 217, 425
\reference Mathis, J.S. \& Whiffen, G. 1989, \apj\ 341, 808.
\reference Mennella, V., Baratta, G.A., Colangeli, L., Palumbo, P., Rotundi, A., Bussoletti, E., \& Strazzulla, G. 1997, \apj\ 481, 545
\reference Mennella, V., Brucato, J.R., Colangeli, L., \& Palumbo, P. 1999, \apj\ 524, L71
\reference Mennella, V., Colangeli, L., Blanco, A., Bussoletti, E., Fonti, S., Palumbo, P., \& Mertins, H.C. 1995, \apj\ 444, 288
\reference Mennella, V., Colangeli, L., Bussoletti, E., Palumbo, P., \& Rotundi, A. 1998, \apj\ 507, L177 
\reference Meyer, D.M. 1997, in Astrophysical Implications of the Laboratory Study of Presolar Materials, ed. T.J. Bernatowicz \& E.K. Zinner, The American Institute of Physics, p. 507
\reference Meyer, D.M, Jura, M., \& Cardelli, J.A. 1998, \apj\ 493, 222
\reference Papoular, R., Conrad, J., Guilois, O., Nenner, I., Reynaud, C., \& Rouzaud, J.-N. 1996, \aap\ 315, 222
\reference Papoular, R., Breton, J., Gensterblum, G., Nenner, I., Papoular, R.J., \& Pireaux, J.-J. 1993, \aap\ 270, L5
\reference Predehl, P., \& Klose, S. 1996, \aap\ 306, 283
\reference Puget, J.L., \& Leger, A. 1989, \araa\ 27, 161
\reference Purcell, E.M., \& Pennypacker, C.R. 1973, \apj\ 186, 705
\reference Rouleau, F., Henning, T., \& Stognienko, R. 1997, \aap\ 322, 633
\reference Rush, W.F., \& Witt, A.N. 1975, \aj\ 80, 31
\reference Schmidt, G.D., Cohen, M., \& Margon, B. 1980, \apj\ 239, L133
\reference Schnaiter, M., Mutschke, H., Dorschner, J., Henning, Th.,\& Salama, F. 1998, \apj\ 498, 486
\reference Seahra, S., \& Duley, W.W. 1999, \apj\ 520, 719
\reference Snow, T.P. \& Witt, A.N. 1996, \apj\ 468, L65
\reference Snow, T.P. \& Witt, A.N. 1995, SCIENCE 270, 1455
\reference Smith, R.K., \& Dwek, E. 1998, \apj\ 503, 831
\reference Sofia, U.J., Cardelli, J.A. Guerin, K.P., \& Meyer, D.M. 1997, \apj\ 482, L10
\reference Soifer, B.T., Houck, J.R., \& Neugebauer, G. 1987, \araa\ 25, 187
\reference Sorrell, W.H. 1990, MNRAS 243, 570
\reference Szomoru, A., \& Guhathakurta, P. 1998, \apj\ 494, L93
\reference Trumpler, R.J. 1930, Lick Obs. Bull., 14, 154
\reference Ugarte, D. 1995, \apj\ 443, L85
\reference van Breda, I.G., \& Whittet, D.C.B. 1981, MNRAS 195, 79
\reference Whittet, D.C.B. 1992, Dust in the Galactic Environment, IOP Pblishing
   Ltd., New York
\reference Witt, A.N., Gordon, K.D., \& Furton, D.G. 1998, \apj\ 501, L111
\reference Witt, A.N., Petersohn, J.K., Holberg, J.B., Murthy, J., Dring, A., \& Henry, R.C. 1993, \apj\ 410, 714
\reference Zinner, E. 1997, Astrophysical Implications of the Laboratory Study of Presolar Materials, ed. T.J. Bernatowicz and E.K. Zinner, (Woodbury: The American Institute of Physics), p. 3
\reference Zubko, V.G., Smith, T.L., \& Witt, A.N. 1999, \apj\ 511, L57
\reference Zubko, V.G., Krelowski, J., \& Wegner, W. 1996, MNRAS 283, 577
\reference Zubko, V.G., Krelowski, J., \& Wegner, W. 1998, MNRAS 294, 548
\reference Zubko, V.G. 1999a, The Physics and Chemistry of the Interstellar Medium, Proc. of the 3rd Cologne-Zermatt Symposium, eds. V. Ossenkopf, J. Stutzki, and G. Winnewisser, GCA-Verlag Herdecke, in press
\reference Zubko, V.G. 1999b, \apj\ 513, L29
\end{references}
\end{document}